\documentclass[10pt,twocolumn,letterpaper]{article}

\usepackage{ijcb}
\usepackage{times}
\usepackage{epsfig}
\usepackage{graphicx}
\usepackage{amsmath}
\usepackage{amssymb}
\usepackage{subcaption}

\ijcbfinalcopy 

\begin{document}

\title{Vocal Style Factorization for Effective Speaker Recognition in \\ Affective Scenarios  }

\author{Morgan Sandler\\
Michigan State University, USA\\
{\tt\small sandle20@msu.edu}
\and
Arun Ross\\
Michigan State University, USA\\
{\tt\small rossarun@cse.msu.edu}
}

\maketitle
\thispagestyle{empty}

\begin{abstract}
The accuracy of automated speaker recognition is negatively impacted by change in emotions in a person's speech. In this paper, we hypothesize that speaker identity is composed of various vocal style factors that may be learned from unlabeled data and re-combined using a neural network to generate a holistic speaker identity representation for affective scenarios. In this regard, we propose the E-Vector architecture,  composed of a 1-D CNN for learning speaker identity features and a vocal style factorization technique for determining vocal styles. Experiments conducted on the MSP-Podcast dataset demonstrate that the proposed architecture improves state-of-the-art speaker recognition accuracy in the affective domain over baseline ECAPA-TDNN speaker recognition models. For instance, the true match rate at a false match rate of 1\% improves from 27.6\% to 46.2\%.  
\end{abstract}

\section{Introduction}
Human speech is composed of various speaking styles that convey conversational semantics through emotions, tone, and other paralinguistic cues. Paralinguistics refer to the non-lexical aspects of speech, such as intonation, pitch, and rhythm, that convey additional meaning beyond the words spoken \cite{prosodycues}. For instance, the tone of voice can make a significant difference in conveying a speaker's conviction or lack thereof, even when the words are the same. In addition, different argumentative styles of speaking may vary based on the speaker’s origin, language, culture, and other factors. These variations in human speech can challenge state-of-the-art speaker recognition systems \cite{pappagari2020x, chowdhury2022domain}. Such systems generally analyze two speech samples and determine whether they correspond to the same individual or not.

In this work, we hypothesize that human speech is a superimposition of multiple vocal style factors and that these factors contribute to both the speaker identity as well as the speaker emotions. By {\em decomposing} an input speech sample to individual vocal style factors and then {\em learning} how to combine them in order to model speaker identity, we alleviate the negative impact of emotions on speaker recognition. Previous state-of-the-art techniques in speaker recognition (also see Table 1) have focused mainly on neutral tones, failing to capture the affective scenario, where emotions modulate the speaking style. While state-of-the-art neural network models can be fine-tuned and adapted to affective scenarios using affective datasets, their accuracy remains underwhelming in many affective scenarios. 

In certain situations, such as those related to national security, it is imperative for an automated speaker recognition to confidently and accurately identify individuals, especially in the affective scenario. This serves as the motivation for our work. In this work, we leverage a voice synthesis technology, referred to as Global Style Tokens (GST) \cite{gst}, to learn vocal style factors and a 1-D CNN to learn identity features directly from raw audio data. The learned vocal style factors are analogous to basis vectors of a style space. We learn these factors from training speech samples in an unsupervised manner while training the speaker recognition system with identity labels. This core method, combined with affective training data and a triplet loss function \cite{ge2e}, ensures that we learn an effective speaker identity representation in the context of affective scenarios.

\section{Learning Emotion Invariant Speaker Identity Features}
Previous studies have examined the relationship between emotion and speaker identity \cite{pappagari2020x, aldeneh2021you, munot2019emotion}. These studies found that the accuracy of speaker recognition is affected by variations in the speaker’s emotional state. The authors of \cite{reliability} proposed a method for estimating the reliability of speaker recognition models based on their emotion content. Emotion recognition models that use adversarial training to separate emotion and speaker identity to achieve speaker-invariance have been proposed \cite{speakerinvariant}. Others methods have been proposed to mitigate affective content and achieve emotion-invariance \cite{nassif2022emotional, simic2022speaker, koolagudi2012speaker}. In contrast to these works, our proposed method uses unsupervised learning of vocal styles, which form the basis of the style space in which we theorize identity is embedded. Instead of learning speaker identity directly from speech samples, we first learn vocal styles and then learn speaker identity as a superimposition of those styles. In the following sections, we will provide more details about our approach to achieving emotion-invariance in speaker recognition.

\subsection{Speech Preprocessing}
A Voice Activity Detector \cite{speechbrain} is used to remove non-speech parts of an input audio. If a sample is longer than 2 seconds, then we split it into many 2-second non-overlapping audio samples. The 2-second speech audio samples are then framed and windowed into multiple smaller audio segments which we refer as speech units. Using a hamming window of length 20 ms and stride 10 ms, each speech unit of 20 ms sampled at 16000 Hz is represented by an audio vector of length 320. The sliding window extracts a speech unit every 10 ms from the 2-second audio sample which extracts approximately 200 speech units per 2-second audio sample. The speech units are then stacked horizontally to form a two-dimensional 320 $\times$ 200 speech representation that we refer to as a speech frame. The extracted speech frames are then organized into triplets during training which are input to the E-Vector model. A single triplet consists of a positive sample, a negative sample, and an anchor sample. The positive and anchor sample share the same speaker identity label whereas the negative and anchor sample do not.

\subsection{Speech Feature Extraction through a 1-D CNN}
We use a 1-D CNN to learn domain-relevant features directly from speech frames rather than from explicitly extracted features. Many popular methods use handcrafted features, such as the mel-frequency cepstrum (MFC) or linear predictive coding (LPC). However, handcrafted features may favor certain domains over others by their inherent design. For example, MFC is used to model the spectral shape of sound, therefore it may not capture enough information to disambiguate between two emotions with similar sounds. Affective feature sets such as the Geneva Minimalistic Acoustic Parameter Set (GeMAPS) have been proposed to address this problem. Recent speaker recognition literature has demonstrated the efficacy in learning features from audio, but none have targeted affective speech directly. A table of popular feature representations and their intended use is shown in Table \ref{tab:handcraftedfeatures}. In this work, we learn a 40-dimensional feature set directly from speech frames. This has been demonstrated previously \cite{deepvox} to retrieve relevant short and long-term behavioral features that are pertinent to speaker identity. 1-D CNN is also beneficial for our deep learning architecture because it can be trained end-to-end along with the GST network (see Section 3) to maximize model learning. The exact architecture setup used is depicted in Figure \ref{fig:evector}.

\begin{table*}[h!]
\caption{An overview of recent speech-based feature representations in speaker recognition.}
\resizebox{\textwidth}{!}{\begin{tabular}{|c|c|c|c|}
\hline
{\underline{ \textbf{Paper}}}                                                                                                                            & {\underline{ \textbf{Feature Category}}} & {\underline{\textbf{Feature Name}}}                         & {\underline{ \textbf{Intended Use}}} \\ \hline
Snyder et al. (2018)    \cite{xvector}                                                                                        & Short-term spectral                       & X-Vector & NN-based Speaker Identity   \\ \hline
Desplanques et al. (2020) \cite{ecapa}                                                                                           & Short-term spectral                       & ECAPA-TDNN & NN-based Speaker Identity  \\ \hline
Chowdhury et al. (2020) \cite{deepvox}                                                                                           & Short-term spectral                       & DeepVOX & NN-based Speaker Identity in Noisy Environments  \\ \hline
Chowdhury et al. (2021) \cite{chowdhDeepTalk21}                                                                                           & Short-term \& Long-term spectral                       & DeepTalk & NN-based Vocal Style Transfer  \\ \hline
Zhang et al. (2022) \cite{zhang2022mfa}                                                                                           & Short-term \& Long-term spectral                       & MFA-Conformer & NN-based Speaker Identity  \\ \hline
Koluguri et al. (2022) \cite{koluguri2022titanet}                                                                                           & Short-term \& Long-term spectral                       & TitaNet & NN-based Speaker Identity  \\ \hline

This work                                                                                            & Short-term \& Long-term Affective spectral                       & E-Vector & NN-based Speaker Identity in Affective Scenarios  \\ \hline

\end{tabular}}
\label{tab:handcraftedfeatures}
\end{table*}

The architecture consists of six 1-D dilated convolution layers \cite{chowdhury2019fusing}, separated by scaled exponential linear unit (SELU) activation functions \cite{selu}. The one-dimensional filters are designed to extract features from individual speech units within a speech frame, rather than across them, based on the assumption that the speaker-dependent characteristics within each unit are independent from those in other units within the same frame. To this end, each 320-dimensional speech unit is processed by a series of 1D dilated convolutional layers, resulting in 40 filter responses that form the short-term spectral representation for that unit. This combination of dilation and scaled activation ensures a larger receptive field, allowing the model to learn sparse relationships between the feature values within a speech unit, which leads to significant performance benefits. Moreover, this method has been shown to be effective in bypassing noise and other small perturbations present in the audio waveform \cite{deepvox}. The larger receptive field facilitates learning of relevant high-dimensional affective information at the feature-extraction level.

\section{Extracting Vocal Style Factors via GSTs}
Text-to-speech is a challenging inverse problem in speech recognition. The goal is to generate human-like speech from a provided text script. There are many challenges that arise when deciding what that voice should sound like. For instance, the system must produce a natural voice with pre-specified vocal style, gender, tone, emotions, and other desirable attributes. Many people convey semantic meaning through their conversations using tone, vocal style, and emotion. Therefore, in text-to-speech, many techniques aim to learn speaker characteristics.

With recent advances in deep-learning, voice synthesis methods have demonstrated success in learning speaker characteristics in order to generate natural-sounding voices. It is for this very reason we used Global Style Tokens (GST), a voice synthesis technique, as our vocal style factor decomposition method.

The GST method was introduced in \cite{gst}. It was proposed for use in a text-to-speech system to encode a vocal style embedding from speech samples. In this method, a voice synthesizer is trained using a concatenated text and vocal style embedding. The vocal style embedding is learned through a multi-head attention module that calculates the similarities between a speaker identity embedding learned from the speech sample and a 10-dimensional bank of embeddings which we will refer to as ``vocal style factors" or VSF. A weighted combination of VSFs produces the style embedding, which is joined with a text embedding, to provide a precise representation of speaker characteristics to a vocoding network.

GSTs have yielded promising results in the emotional voice conversion (EVC) literature \cite{kunzhoumixed} and have demonstrated the ability to encode pitch contours that are vital to vocal style in human speech \cite{chowdhDeepTalk21}. Therefore, we incorporate a similar approach into our E-Vector architecture. In our E-Vector model, we swap reconstruction loss with generalized end-to-end loss \cite{ge2e}. This ensures that we will learn speaker identity in the context of distinguishing genuine and impostor pairs for speaker verification. 

\section{Proposed E-Vector Architecture}

\begin{figure*}[htb!]
    \centering
    \includegraphics[scale=0.7]{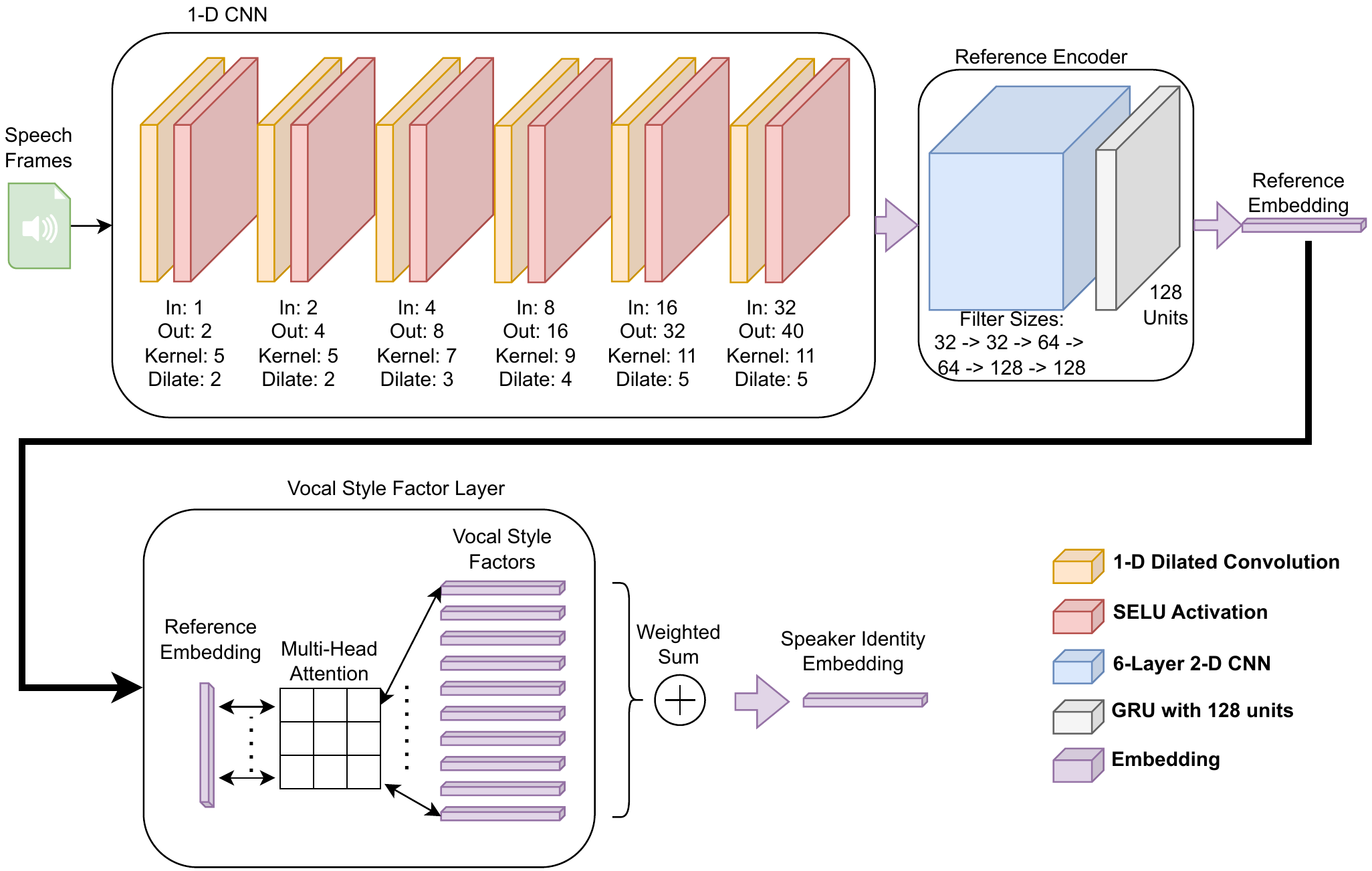}
    \caption{The proposed E-Vector architecture. The 1-D CNN takes speech frames (detailed in Section 2.1) as input and outputs a vector of 40 features. These features are then passed to a reference encoder, which extracts a fixed-dimension embedding representing the speaker identity features. The vocal style factor layer decomposes the identity embedding into a fixed number of factors (e.g., 10 factors) via multi-head attention (8 heads). Finally, the weighted combination of these factors yields a speaker identity embedding.}
    \label{fig:evector}
\end{figure*}

\begin{figure}
  \centering
  \includegraphics[width=.9\linewidth]{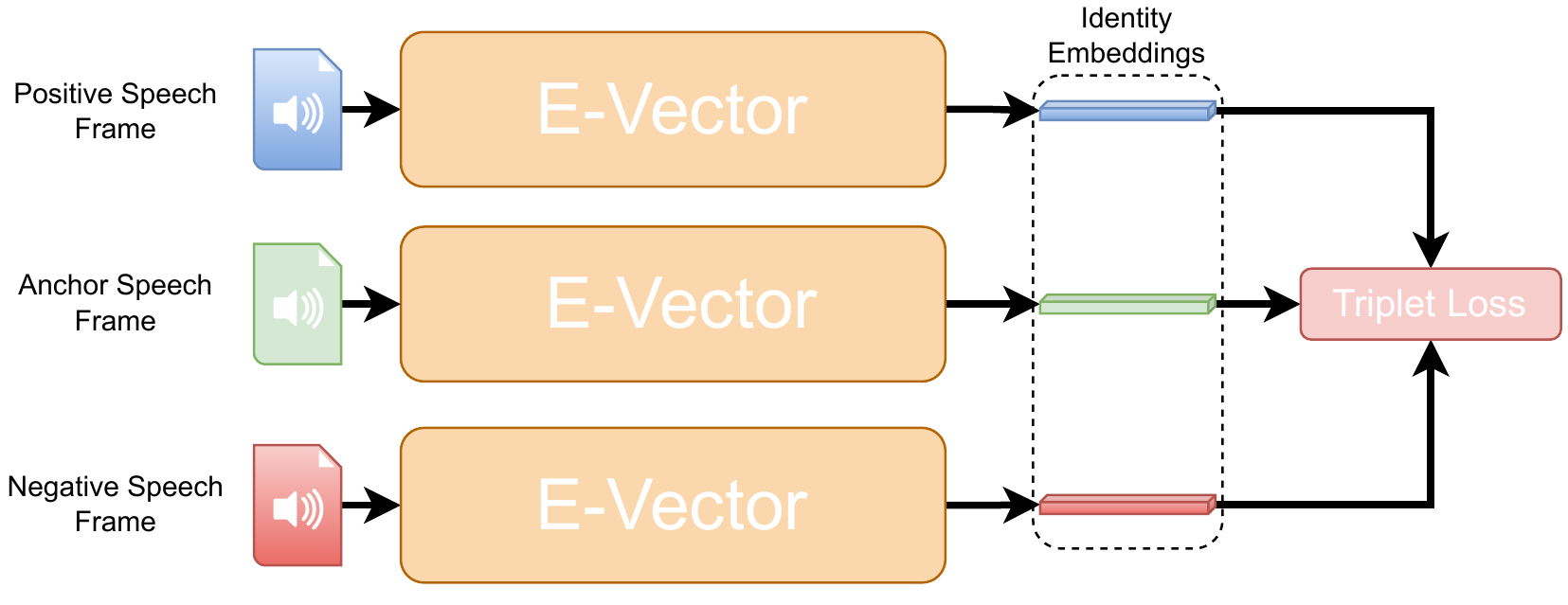}
  \captionof{figure}{E-Vector \textbf{Training} Setup.}
  \label{fig:trainingsetup}
\end{figure}
\begin{figure}
  \centering
  \includegraphics[width=.9\linewidth]{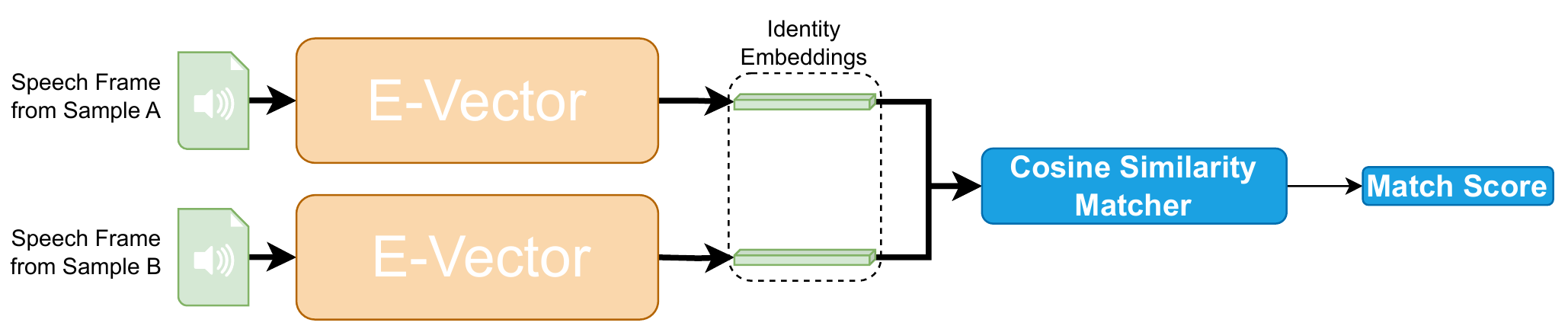}
  \captionof{figure}{E-Vector \textbf{Testing} Setup.}
  \label{fig:testingsetup}
\end{figure}

We propose E-Vector as a supervised learning model designed for speaker recognition in varying affective scenarios. E-Vector consists of two networks trained end-to-end. The first network is a 1-D CNN which learns a 40-dimensional feature set directly from speech frames, while a GST network, which consists of a 2-Dimensional CNN and Gated Recurrent Unit, produces 128-dimensional reference embeddings. It does so by learning 10 vocal style factor embeddings that serve as basis vectors for a learned vocal style space that encapsulates the affective styles inherent in the data. The final 256-dimensional speaker representation learned is referred to as an E-Vector. The E-Vector architecture is illustrated in Figure \ref{fig:evector}. The training and testing of the model are illustrated in Figures \ref{fig:trainingsetup} and \ref{fig:testingsetup}, respectively.

\section{Loss Functions}
{\bf Additive Angular Margin Loss (AAM)}, often referred to as ArcFace loss \cite{arcface}, is a popular loss function proposed to learn geodesic distances between identity embeddings on a hypersphere. The margin hyperparameter enforces compactness between genuine pairs and more distance between impostor pairs to a higher degree than traditional softmax loss. The formulation is as follows: 

\begin{equation}
    L = -\frac{1}{N}\sum_{i=1}^{N}\log\frac{e^{s(\cos(\theta_{y_i}+m))}}{e^{s(\cos(\theta_{y_i}+m))} + \sum_{j=1, j \neq y_i}^{n}e^{s\cos\theta_j}}
\end{equation} 

Here, $N$ is the size of the batch, $n$ denotes the number of identities, $s$ is the radius of the hypersphere, $m$ is the additive angular margin penalty between the feature and the weight vectors, and $\theta$ is the angle between the weight and feature vectors.

{\bf Generalized End-to-End Loss (GE2E)} \cite{ge2e} is specifically derived for speaker verification and is designed to penalize more those impostor pairs that may frequently false-match. It does this by identifying the most similar-to-genuine false match speakers as its basis for maximizing the distance between genuine embeddings and impostor embeddings. The genuine term in GE2E loss corresponds to a genuine match between the input embedding vector and its true speaker label. Conversely, the impostor term corresponds to an impostor match between the input embedding vector and the speaker label with the highest similarity to genuine among all false speakers. In prior literature, GE2E loss has demonstrated effectiveness and does not require an initial example selection stage as needed by loss functions such as Tuple End-To-End Loss \cite{ge2e}. The importance in identifying hard speaker verification samples, which is common in affective speaker recognition, is why we use this loss in our work.

\begin{equation}
    L = \sum_{i=1}^N \left( \alpha_i d(x_i, y_i)^2 + (1 - \alpha_i) \max_{j \neq i} d(x_i, y_j)^2 \right)
\end{equation}

Here, $x_i$ is the genuine embedding for utterance $i$, $y_i$ is the ground truth embedding for utterance $i$, $\alpha_i$ is the weight assigned to each utterance, $d(x_i, y_i)$ is the distance between a genuine pair, and $d(x_i, y_j)$ is the distance between the genuine embedding $x_i$ and the impostor embedding $y_j$.

\section{MSP-Podcast Dataset}
MSP-Podcast \cite{msp} is a natural emotion dataset derived from podcasts and labeled through crowd-sourced consensus voting. Each sample is labeled with a discrete emotion (e.g., `Happy') as well as continuous emotion dimensions (valence, arousal, dominance) using a 5-point Likert scale. There are 10 different labels for discrete emotions while continuous emotion values range between 0 and 5.

The MSP-Podcast dataset contains a total of 73,042 speaking turns, equivalent to 110 hours of speech. The dataset is divided into several sets for training and evaluation purposes. Test set 1 contains segments from 60 speakers (30 female and 30 male), totaling 15,326 segments. Test set 2 consists of 5,037 segments randomly selected from 100 podcasts. These segments are not included in any other partition. The Validation set includes segments from 44 speakers (22 female and 22 male), totaling 7,800 segments. \textbf{Note, we use the validation set as another testing set, and its data is not seen during the training process}. The Train set contains the remaining speech samples, totaling 38,179 segments.

In this work, we use the speaker identity labels for our training process. For evaluation purposes, we use the Test 1, Test 2, and Validation sets---each with its own unique composition---to evaluate our proposed E-Vector model.

\section{Experiment Setup}
\subsection{Models and Baselines}
Four models are trained for the speaker verification task: ECAPA-TDNN/Vox (denoted as E1), ECAPA-TDNN/MSP (denoted as E2), ECAPA-TDNN/Vox+MSP (denoted as E3), and E-Vector (denoted as E4). E1, E2, and E3 serve as the baseline models to evaluate different training configurations of state-of-the-art speaker recognition systems to determine their efficacy in the affective scenario. 
E1 is a popular state-of-the-art ECAPA-TDNN \cite{ecapa} speaker recognition model pre-trained on VoxCeleb 1 \& 2 datasets. E2 also uses the ECAPA-TDNN architecture but we substitute the AAM loss with the GE2E loss; further, the network is trained from scratch using the MSP-Podcast train set. Our third baseline, E3, uses the pre-trained weights from E1, but we fine-tune the last layer (1.5M parameters) using the MSP-Podcast train set. E4 is the proposed E-Vector approach and is trained with the MSP-Podcast train set. The specifics of training the four models are detailed in the next section.

\subsection{Training Details}
All four models are trained for the speaker verification task, and training is stopped if there are no empirically significant decreases to the train loss over a window of 20,000 steps. All copies of trained models and their respective training/evaluation codes are available via email request or on our Github website. All models were trained over 5 days with an Nvidia RTX 2080 Ti GPU.

The {\bf ECAPA-TDNN/Vox (E1)} pre-trained weights are accessed from HuggingFace \cite{huggingface}, while the training data is sourced from the VoxCeleb1 and VoxCeleb2 datasets. The authors trained this model for 12 epochs and fit 22.2M trainable parameters using an AAM loss function. No MSP-Podcast train data is used in the E1 training process. This model serves as an off-the-shelf speaker verification model comparison to E-Vector.

The {\bf ECAPA-TDNN/MSP (E2)} model is of the ECAPA-TDNN architecture, but the training process uses the GE2E loss function (same as E-Vector) rather than AAM loss. Using the data from the MSP-Podcast train set, 22.2M parameters are learned over 7,500 steps (where overfitting is observed). Overfitting is to be expected as the parameters of the model are large in relation to the quantity of data available. Even with this limitation, we find that this model still achieves competitive results in our affective scenario.

The {\bf ECAPA-TDNN/Vox+MSP (E3)} model uses the same pre-trained weights from E1. To adapt the model to the target domain, we fine-tuned the last layer (1.5M trainable parameters) using the MSP-Podcast training set. E3 employs the AAM loss function. This model represents a popular transfer-learning method in order to determine whether we can address the affective scenario by adapting pre-existing models.

Our proposed {\bf E-Vector (E4)} method uses 10 vocal style factors to train for 105,000 steps, using the GE2E loss function and approximately 959.8K trainable parameters. The MSP-Podcast training set serves as the data source.

\subsection{Model Evaluation}
To evaluate our 3 baseline models and E-Vector, we compute the following metrics: Equal Error Rate (EER), Minimum Detection Cost Function (minDCF), D-Prime (d'), Area Under Curve (AUC), True Match Rate at a specified False Match Rate (TMR @ FMR 1\% and 10\%), Detection Error Tradeoff (DET) curves, and Match Score Distributions. 

\subsubsection{minDCF Metric Formulation}
The minDCF metric allows us to study the costs associated with the speaker recognition system. The formulation is as follows:

{
\scriptsize
\begin{multline}
    DCF(\tau) = C_{miss}P_{miss}(\tau)P_{target} + C_{FM}P_{FM}(\tau)(1-P_{target})
\end{multline}
}
where, $\tau$ is a match decision threshold, $C_{miss}$ is the cost of a missed verification error, $C_{FM}$ is the cost of a false match error, $P_{target}$ is the prior probability of the target speaker occurring in the data, $P_{miss}(\tau)$ is the probability of a miss verification at a given threshold, and $P_{FM}(\tau)$ is the probability of a false match at a given threshold. minDCF is the minimization of this cost function. We evaluate minDCF at $C_{miss}=10$, $C_{FM}=1$. This is because in forensic/security scenarios, it is preferred to identify/match a speaker than to completely miss a potential match at all.

\subsubsection{Speaker Verification Testing Details}
In the MSP-Podcast dataset, there are 109.8M speaker verification pairs in test set 1, 10.7M pairs in test set 2, and 28.8M pairs in the validation set. Due to the large quantity of pairs available, we store all results and evaluate metrics on a smaller, uniformly sampled subset of the results. Computing the full speaker verification tests takes 27 hours on test set 1, 6 hours on test set 2, and 16 hours on the validation set. All experiments were conducted on a Nvidia GeForce 2080 Ti GPU. The approximate size of the reduced evaluation sets is 548K pairs for test set 1 (approx. 0.5\% of computed match scores), 107K pairs for test set 2 (approx. 1\% of computed match scores), and 287K pairs for the validation set (approx. 1\% of computed match scores). All speaker verification results are publicly available by email request or on our Github. Attributes available in the results table include emotion of sample A, emotion of sample B, identity of sample A, identity of sample B, speaker verification match score, continuous emotion dimensions of sample A and sample B, and gender of sample A and sample B. The average inference time of the proposed method is 18 ms per sample.

\section{Results}

 \begin{figure*}[t]
  \centering
  \begin{minipage}{.315\linewidth}
    \centering
    \subcaptionbox{\footnotesize E-Vector Test Set 1 \label{fig:1a} }
    {\includegraphics[width=\textwidth, height=1in]{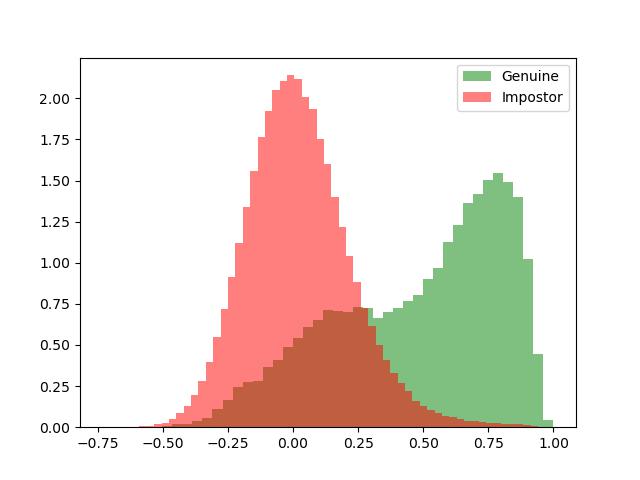}}
  \end{minipage}\quad
  \begin{minipage}{.315\linewidth}
    \centering
    \subcaptionbox{\footnotesize E-Vector Test Set 2 \label{fig:1b}}
    {\includegraphics[width=\textwidth, height=1in]{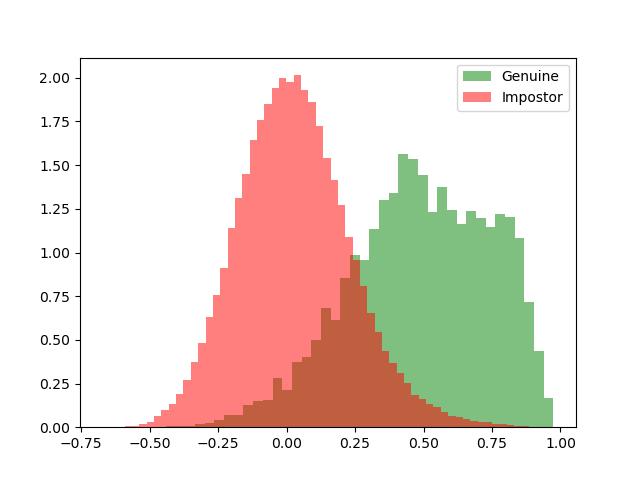}}
  \end{minipage}\quad
  \begin{minipage}{.315\linewidth}
    \centering
    \subcaptionbox{\footnotesize E-Vector Validation Set \label{fig:2}}
    {\includegraphics[width=\textwidth, height=1in]{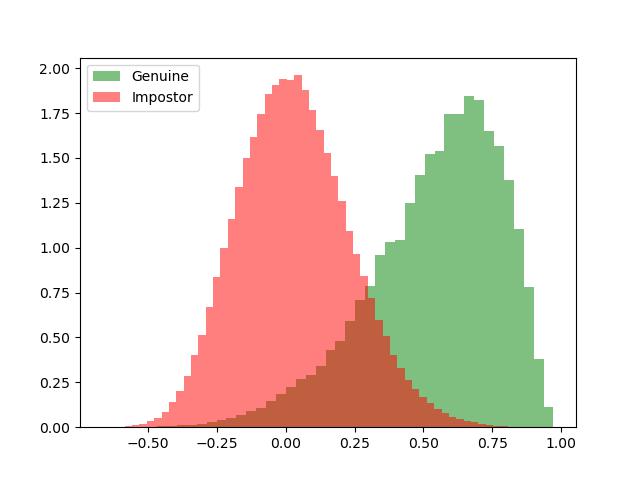}}
  \end{minipage}\quad
    \begin{minipage}{.315\linewidth}
    \centering
    \subcaptionbox{\footnotesize ECAPA-TDNN/Vox Test Set 1 \label{fig:1a} }
    {\includegraphics[width=\textwidth, height=1in]{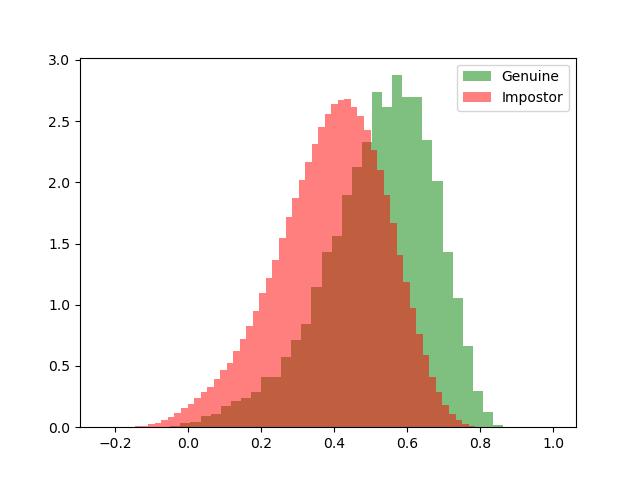}}
  \end{minipage}\quad
  \begin{minipage}{.315\linewidth}
    \centering
    \subcaptionbox{\footnotesize ECAPA-TDNN/Vox Test Set 2 \label{fig:1b}}
    {\includegraphics[width=\textwidth, height=1in]{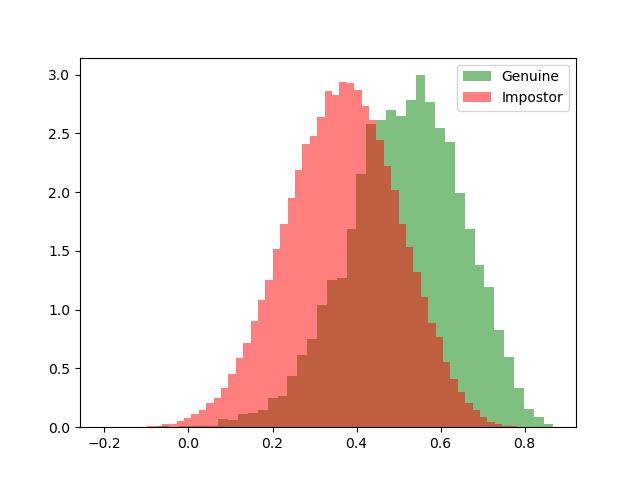}}
  \end{minipage}\quad
  \begin{minipage}{.315\linewidth}
    \centering
    \subcaptionbox{\footnotesize ECAPA-TDNN/Vox Validation Set \label{fig:2}}
    {\includegraphics[width=\textwidth, height=1in]{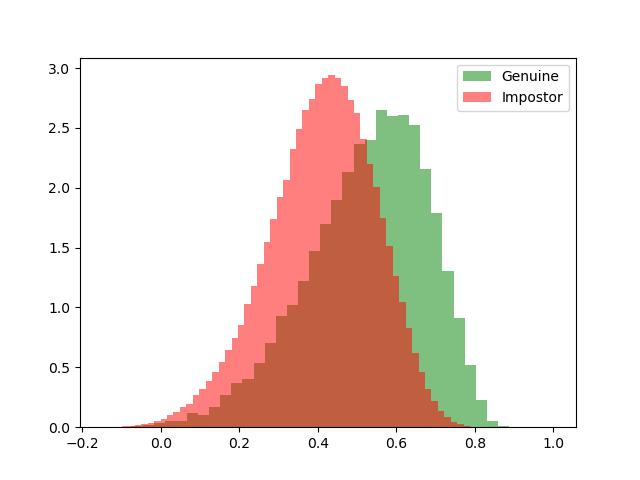}}
  \end{minipage}\quad
      \begin{minipage}{.315\linewidth}
    \centering
    \subcaptionbox{\footnotesize ECAPA-TDNN/MSP Test Set 1 \label{fig:1a} }
    {\includegraphics[width=\textwidth, height=1in]{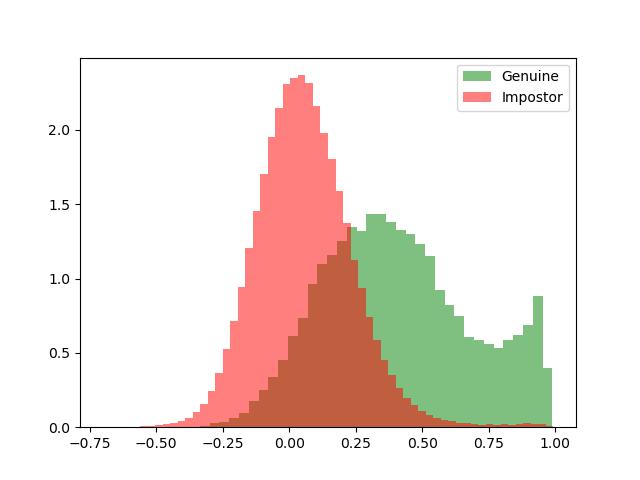}}
  \end{minipage}\quad
  \begin{minipage}{.315\linewidth}
    \centering
    \subcaptionbox{\footnotesize ECAPA-TDNN/MSP Test Set 2 \label{fig:1b}}
    {\includegraphics[width=\textwidth, height=1in]{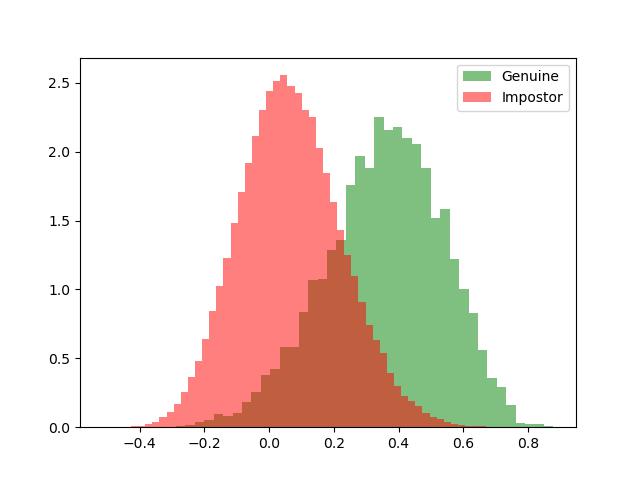}}
  \end{minipage}\quad
  \begin{minipage}{.315\linewidth}
    \centering
    \subcaptionbox{\footnotesize ECAPA-TDNN/MSP Validation Set \label{fig:2}}
    {\includegraphics[width=\textwidth, height=1in]{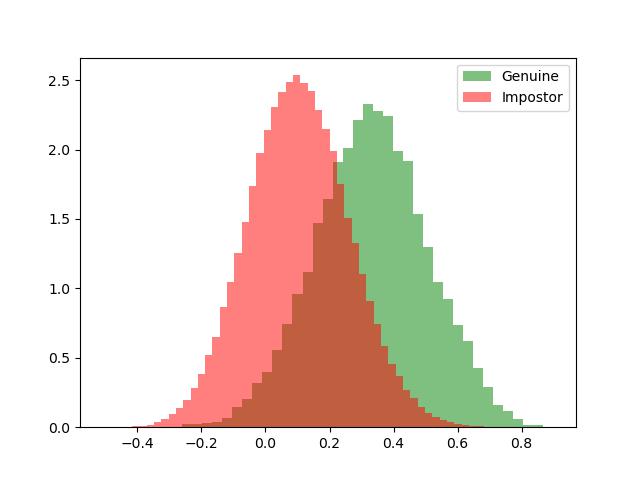}}
  \end{minipage}\quad
        \begin{minipage}{.315\linewidth}
    \centering
    \subcaptionbox{\footnotesize ECAPA-TDNN/Vox+MSP Test Set 1 \label{fig:1a} }
    {\includegraphics[width=\textwidth, height=1in]{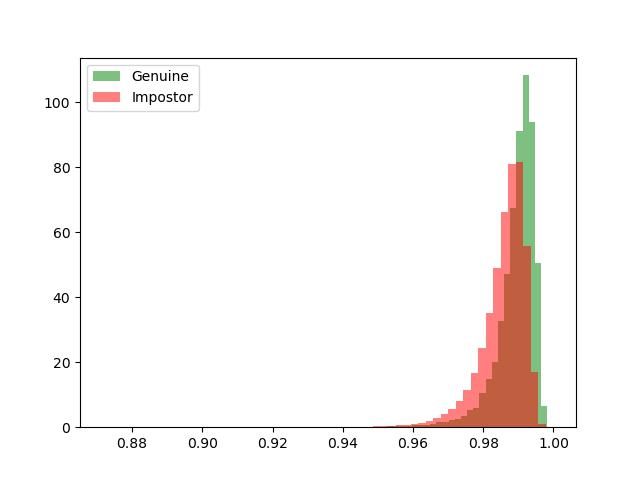}}
  \end{minipage}\quad
  \begin{minipage}{.315\linewidth}
    \centering
    \subcaptionbox{\footnotesize ECAPA-TDNN/Vox+MSP Test Set 2 \label{fig:1b}}
    {\includegraphics[width=\textwidth, height=1in]{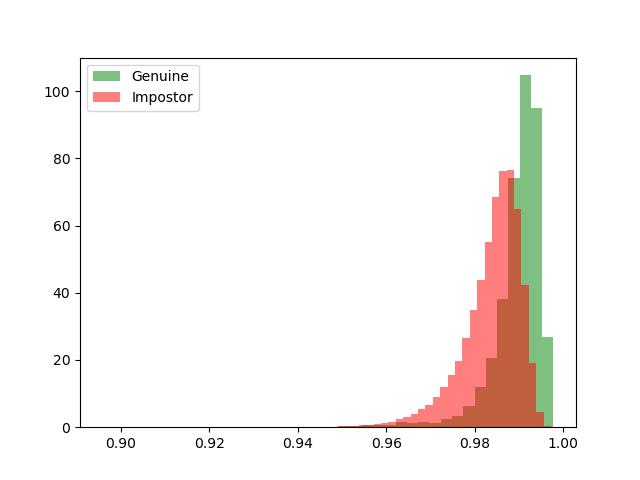}}
  \end{minipage}\quad
  \begin{minipage}{.315\linewidth}
    \centering
    \subcaptionbox{\footnotesize ECAPA-TDNN/Vox+MSP Validation Set \label{fig:2}}
    {\includegraphics[width=\textwidth, height=1in]{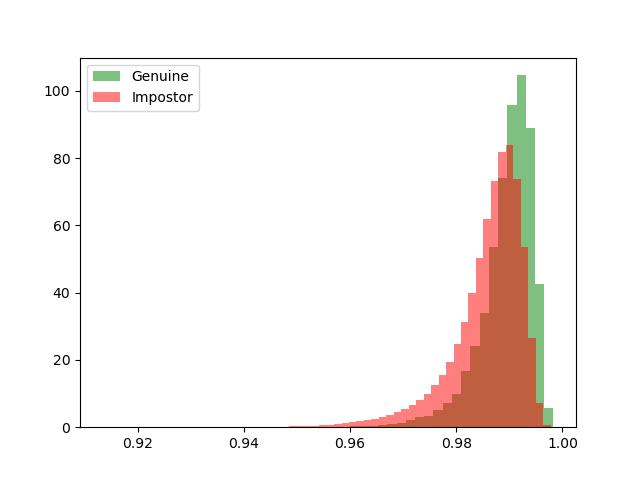}}
  \end{minipage}\quad
  
  \caption{Match score distributions for E-Vector and ECAPA-TDNN baselines on three test sets of MSP-Podcast: test set 1, test set 2, validation set. Note that the validation set was {\em not} used in the training stage. The y-axis denotes density. The x-axis denotes match score.}
  \label{fig:matchscores}
\end{figure*}

\begin{figure*}[h]
  \centering
  \begin{minipage}{.33\textwidth}
    \centering
    \subcaptionbox{\footnotesize Test Set 1 \label{fig:1a}}
    {\includegraphics[width=\textwidth]{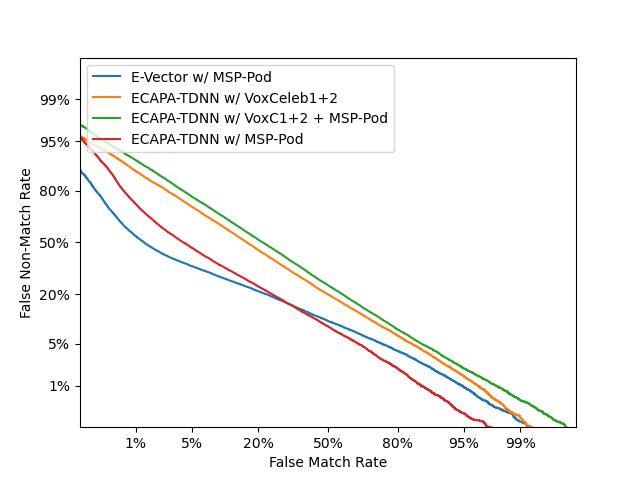}}
  \end{minipage}%
  \begin{minipage}{.33\textwidth}
    \centering
    \subcaptionbox{\footnotesize Test Set 2 \label{fig:1b}}
    {\includegraphics[width=\textwidth]{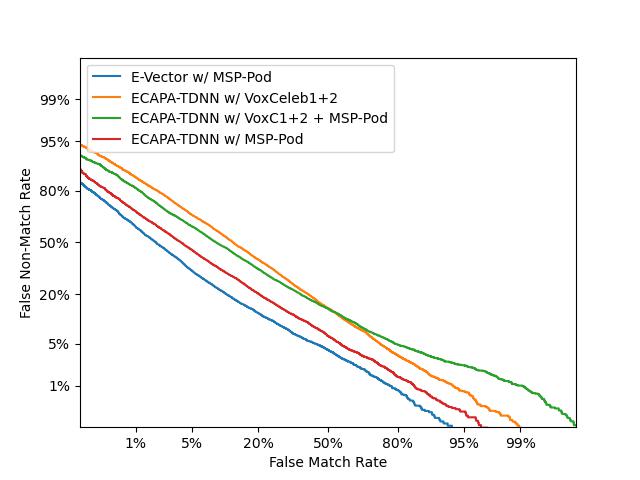}}
  \end{minipage}%
  \begin{minipage}{.33\textwidth}
    \centering
    \subcaptionbox{\footnotesize Validation Set \label{fig:2}}
    {\includegraphics[width=\textwidth]{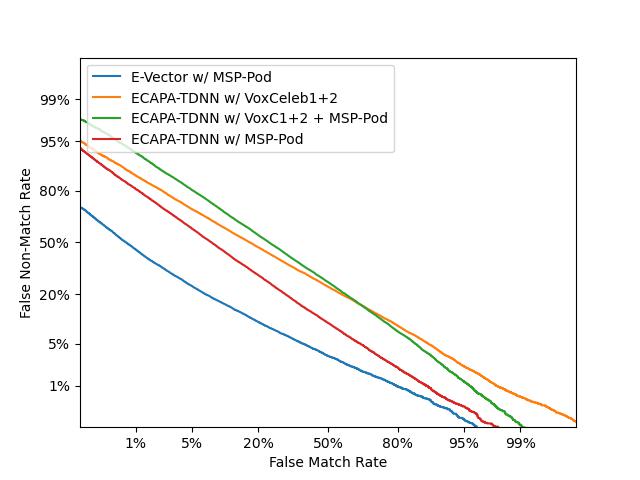}}
  \end{minipage}
  \caption{Detection Error Trade-off (DET) curves comparing E-Vector with the baseline experiments.}
  \label{fig:main}
\end{figure*}


\begin{table*}[t]
\centering
\begin{tabular}{|l|l|l|l|l|l|l|l|}
\hline
\textbf{Model} & \textbf{Trainable Params} & \textbf{Test Set} & \textbf{EER} & \textbf{minDCF} & \textbf{TMR@FMR = \{1\%, 10\%\}} & \textbf{D'} & \textbf{AUC} \\ \hline
\textbf{E1} & 22.2M & Val & 0.34 & 0.094 & 14.7\%, 40.7\% & 0.83 & 0.72 \\ \cline{3-8} 
 &  & Test 1 & 0.33 & 0.096 & 12.0\%, 39.9\% & 0.90 & 0.74 \\ \cline{3-8} 
 &  & Test 2 & 0.29 & 0.094 & 14.0\%, 44.2\% & 1.12 & 0.78 \\ \hline
\textbf{E2} & 1.5M & Val & 0.37 & 0.099 & 7.15\%, 30.0\% & 0.68 & 0.68 \\ \cline{3-8} 
 &  & Test 1 & 0.35 & 0.098 & 8.92\%, 33.7\% & 0.73 & 0.70 \\ \cline{3-8} 
 &  & Test 2 & 0.27 & 0.089 & 20.7\%, 52.5\% & 1.18 & 0.80 \\ \hline
\textbf{E3} & 22.2M & Val & 0.25 & 0.090 & 19.4\%, 55.1\% & 1.35 & 0.83 \\ \cline{3-8} 
 &  & Test 1 & 0.22 & 0.080 & 27.6\%, 65.0\% & 1.53 & 0.86 \\ \cline{3-8} 
 &  & Test 2 & 0.19 & 0.078 & 30.5\%, 68.7\% & 1.68 & 0.88 \\ \hline
\textbf{E4} & 959.8K & Val & \textbf{0.13} & \textbf{0.054} & \textbf{55.5\%, 84.0\%} & \textbf{2.14} & \textbf{0.94} \\ \cline{3-8} 
 &  & Test 1 & \textbf{0.20} & \textbf{0.061} & \textbf{46.2\%, 72.0\%} & \textbf{1.58} & \textbf{0.87} \\ \cline{3-8} 
 &  & Test 2 & \textbf{0.15} & \textbf{0.068} & \textbf{38.7\%, 77.1\%} & \textbf{1.95} & \textbf{0.91} \\ \hline
\end{tabular}
\caption{E-Vector comparison to baseline models. E4 (E-Vector) performs best across all metrics in our study across three separate test sets.}
\label{tab:main_results}
\end{table*}

We note that E-Vector learns certain categories of affective states. This is reflected in the bimodal genuine distribution which can be observed in the E-Vector test set 1 plot in Figure \ref{fig:matchscores}. The bimodal distribution may be a product of overcoming challenges caused by emotion modulation. That may be why we observe a similar, but less pronounced bimodal distribution in the corresponding test set 1 genuine distribution in the ECAPA-TDNN/MSP experiment. We also conclude that fine-tuning ECAPA-TDNN with MSP-Podcast train data does not work as intended, as it finds a niche local optima, perhaps due to an insufficient number of samples. Also, E-Vector improves recognition EER from 0.22 in E3 to 0.20, and TMR@FMR1\% accuracy from 27.6\% in E3 to 46.2\%.

\section{Impact of Vocal Style Factors}
In this analysis, we perform three experiments on the E-Vector architecture with varying numbers of vocal style factors to determine the impact of the number of factors on speaker recognition. We chose 5, 10, and 20 factors and trained each model for 105,000 steps with the MSP-Podcast train set data. The 5-factor model is denoted E5, 10-factor is denoted E6, and 20-factor is denoted E7. Each model takes ~47 hours to train on an Nvidia RTX 2080 Ti. The results are shown in Table \ref{tab:factorsize_metrics}. Generally, we find that the TMR at FMR 1\% improves across all test sets with the addition of more factors. Since speaker identity in our model is composed of many discrete vocal style factors and there may be potentially an indefinite number of speaking styles, it is not practical to have one factor for every possible style. Even with the limited range of vocal style factors tested, we observe a significant increase in performance compared to the baseline ECAPA-TDNN models. That said, it is not clear that more factors are always better and variation may exist across cultures, dialects, etc. Therefore, additional study into this hyperparameter is necessary.

\begin{table*}[t]
\centering
\begin{tabular}{|l|l|l|l|l|l|l|l|}
\hline
\textbf{Model} & \textbf{Number of VSFs} & \textbf{Test Set} & \textbf{EER} & \textbf{minDCF} & \textbf{TMR@FMR = \{1\%, 10\%\}} & \textbf{D'} & \textbf{AUC} \\ \hline
\textbf{E5} & 5 & Val & 0.13 & 0.049 & 60.0\%, 84.9\% & 2.19 & 0.94 \\ \cline{3-8} 
 &  & Test 1 & 0.22 & 0.062 & 46.4\%, 70.5\% & 1.61 & 0.87 \\ \cline{3-8} 
 &  & Test 2 & 0.16 & \textbf{0.068} & 39.8\%, 76.7\% & 1.93 & 0.91 \\ \hline
\textbf{E6} & 10 & Val & 0.13 & 0.054 & 55.5\%, 84.0\% & 2.14 & 0.94 \\ \cline{3-8} 
 &  & Test 1 & \textbf{0.20} & \textbf{0.061} & 46.2\%, \textbf{72.0\%} & 1.58 & 0.87 \\ \cline{3-8} 
 &  & Test 2 & \textbf{0.15} & \textbf{0.068} & 38.7\%, \textbf{77.1\%} & \textbf{1.95} & 0.91 \\ \hline
\textbf{E7} & 20 & Val & \textbf{0.12} & \textbf{0.048} & \textbf{60.8\%, 86.2\%} & \textbf{2.21} & 0.94 \\ \cline{3-8} 
 &  & Test 1 & 0.21 & 0.062 & \textbf{46.9\%}, 70.0\% & \textbf{1.62} & 0.87 \\ \cline{3-8} 
 &  & Test 2 & 0.17 & \textbf{0.068} & \textbf{40.8\%}, 76.1\% & 1.89 & 0.91 \\ \hline
\end{tabular}
\caption{Comparison across three trained E-Vector models with 5, 10, 20 VSFs. Each model is trained to 105,000 steps with the MSP-Podcast train set data.}
\label{tab:factorsize_metrics}
\end{table*}


\section{Impact of Emotion on Match Scores}
\begin{figure}[h!]
    \centering
    \begin{subfigure}{0.24\textwidth}
        \centering
        \includegraphics[width=1\linewidth]{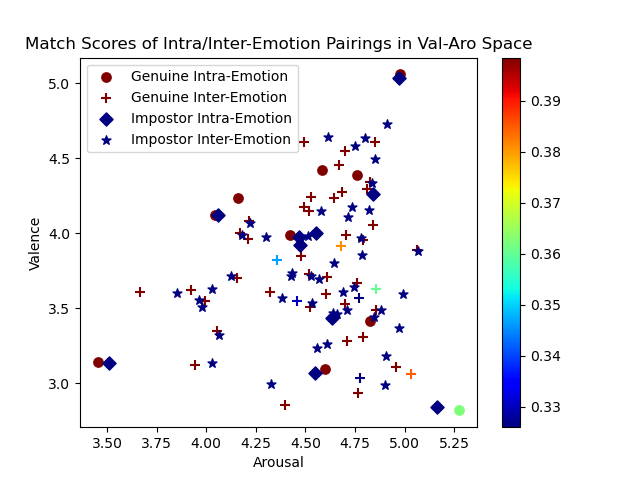}
        \caption{\footnotesize E-Vector}
        \label{fig:1aa}
    \end{subfigure}%
    \begin{subfigure}{0.24\textwidth}
        \centering
        \includegraphics[width=1\linewidth]{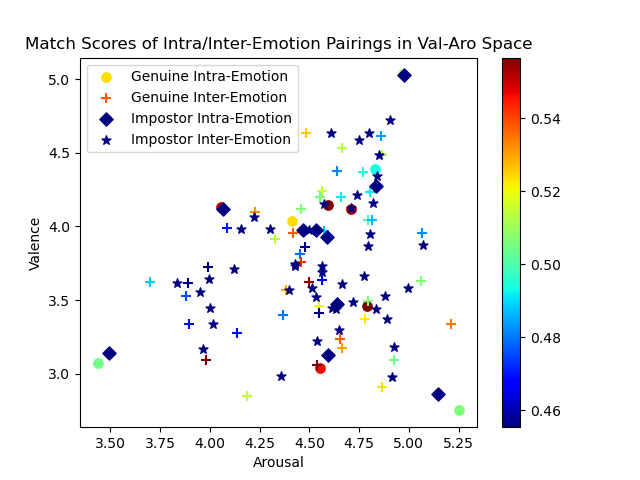}
        \caption{\footnotesize ECAPA-TDNN/Vox}
        \label{fig:2aa}
    \end{subfigure}\\ 
    \begin{subfigure}{0.24\textwidth}
        \centering
        \includegraphics[width=1\linewidth]{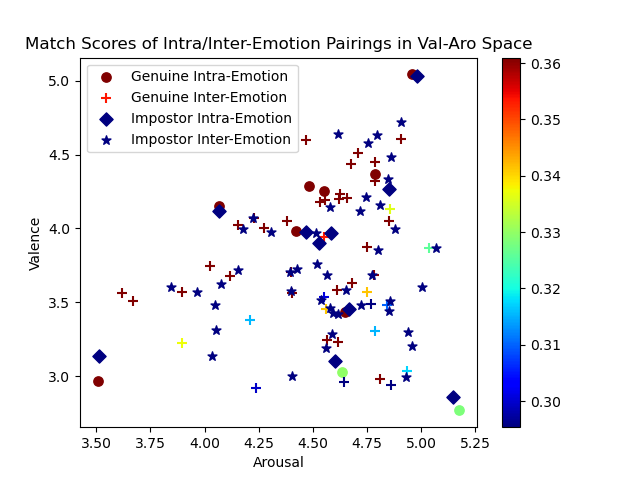}
        \caption{\footnotesize ECAPA-TDNN/MSP}
        \label{fig:3aa}
    \end{subfigure}%
    \begin{subfigure}{0.24\textwidth}
        \centering
        \includegraphics[width=1\linewidth]{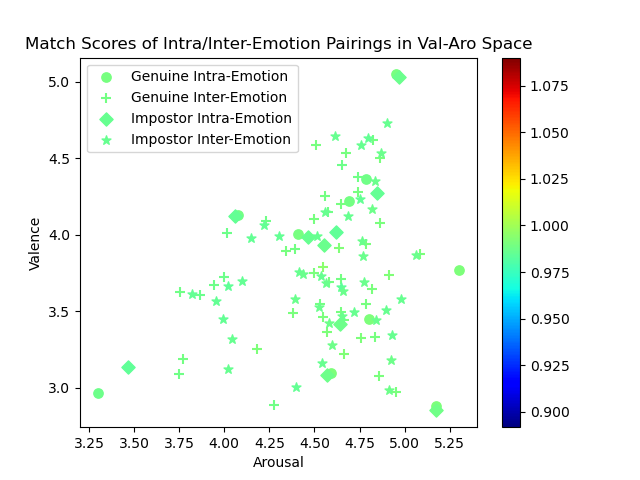}
        \caption{\footnotesize ECAPA-TDNN/Vox+MSP}
        \label{fig:4aa}
    \end{subfigure}
    \caption{Inter/Intra-emotion speaker recognition test set 1 match scores visualized in the valence-arousal emotion space.}
    \label{fig:valaro}
\end{figure}

In an ideal setting for speaker recognition systems in an affective scenario, the scores across different emotions (inter-emotion) and within same emotions (intra-emotion) should closely group together. Additionally, there should be a clear boundary that distinguishes genuine scores from impostor scores.
To analyze our proposed method, we represent the average scores between each emotion pair (e.g., happy-anger) for both genuine and impostor classes. This involves 90 inter-emotion  scores and 20 intra-emotion scores, totaling 110 match score averages. We present these scores in Figure \ref{fig:valaro}, within the frame of the continuous emotion dimension space, characterized by arousal and valence.
In the perfect scenario, this plot would have inter/intra-emotion genuine scores shown in one color, while all the inter/intra-emotion impostor match scores would be in a different color. The proposed method shows a tendency to decrease the differences between intra- and inter-emotion match scores for both the genuine and impostor classes,  compared to the baseline methods.
These findings correspond with the overall performance metrics that we observe in Table \ref{tab:main_results}. While the E-Vector model does  improve overall performance, it is still not clear which continuous emotion dimensions, and their associated emotion categories, have the most impact on match scores. This topic necessitates further study.

\section{Summary and Future Work}

Learning speaker identity in voice has traditionally been a 1-step process—training a model to learn speaker identity from speech samples. This has effectively worked in most {\em neutral}  speech scenarios, but in the dynamic, {\em affective} scenario, the performance sharply degrades. In this work, we propose a 2-step process: first, learning global styles of speech patterns based on thousands of speaker identities, and second, representing speaker identity as combinations of those learned vocal style patterns.

The E-Vector model architecture incorporates this 2-step view of the speaker identity. The first step learns similarities (via multi-head attention) between thousands of people’s speaking patterns and creates embeddings of those vocal styles. The advantage of the architecture is that it is not required to label those speaking patterns, although the patterns should be salient in the training data. This enables E-Vector to learn vocal patterns that we perhaps have not yet discovered or considered. While it may be capable of learning new patterns from data, this functionality necessitates the use of high-quality data to operate effectively. Despite this constraint, we chose to utilize the MSP-Podcast dataset, which includes preprocessing steps designed to minimize the presence of low-quality data during the collection phase \cite{msp}.

Then, in the second step, the E-Vector model learns speaker identity, but only as a weighted combination of the aforementioned vocal styles.

By modeling identity as a composition of learned vocal style factors, we find that our proposed method, E-Vector, outperforms state-of-the-art ECAPA-TDNN baseline models on the task of speaker recognition in affective scenarios where emotions have an impact on the speaking style. In addition, we explore the relationship between the number of vocal style factors used in training and the eventual performance. With 20 VSFs in the E-Vector architecture, we are able to obtain a TMR of 46.9\% @ FMR of 1\% on MSP-Podcast test-set-1, which is 19.3\% higher than the best ECAPA-TDNN baseline. 

In recent literature, there has been a focus on improving speaker recognition models by learning the weights using multiple modalities such as text and audio samples, and then using only audio during the inference stage for identity assessment \cite{nist2021}. Therefore, in the affective domain, further work is necessary to obtain larger speech datasets with associated text transcripts to harness these novel capabilities. 

In addition to utilizing more modalities in the training process, a semi-supervised training scheme may be employed. Currently, the E-Vector architecture does not use any direct supervision via emotion labels to tune its weights; in other words, it is tuned by the GE2E loss. Future work could include an “emotion” loss in addition to the GE2E loss. The rationale would be for the model to predict an emotion label, and the corresponding true emotion labels would adjust the loss and weights of the vocal style factors accordingly to disentangle emotion and identity further.

The E-Vector architecture, while originally designed for affective scenarios, is not confined to processing affective data. Further assessment of the E-Vector model on standard speaker recognition benchmarks like VoxCeleb and Librispeech, coupled with comparisons to newer speaker recognition architectures such as TitaNet or MFA-Conformer, could provide valuable insights into its potential for generalization.

\section{Reproducibility}
All training, evaluation, and analysis code and experiment results will be made available on our Github link.\footnote{Training/Evaluation Code: https://github.com/morganlee123/evector} Trained models and embeddings have large file sizes and will be made available upon request.

\clearpage
{\small
\bibliographystyle{ieee}
\bibliography{main}
}

\end{document}